\documentclass[conference]{IEEEtran}
\IEEEoverridecommandlockouts
\usepackage{cite}
\usepackage{amsmath,amssymb,amsfonts}
\usepackage{algorithmic}
\usepackage{graphicx}
\usepackage{textcomp}
\usepackage{xcolor}

\usepackage{stackengine}
\usepackage[textwidth=1.7cm, disable]{todonotes}
\usepackage{subfigure}
\usepackage{cleveref}
\usepackage{booktabs}
\usepackage{xspace}
\usepackage{balance}
\usepackage{array}
\usepackage{multirow}
\usepackage{bbm}
\usepackage{url}

\usepackage{colortbl}
\newtheorem{example}{Example}[section]
\usepackage{pgfplots}

\renewcommand{\phi}{\varphi}

\newcommand{\marginnote}[3]{\todo[color=#3!40,size=\footnotesize]{\textbf{#2:} #1}}

\newcommand{\mgmargin}[1]{\marginnote{#1}{MG}{yellow}}
\newcommand{\mgmargindone}[1]{}

\newcommand{\mginlinedone}[1]{}

\newcommand{\bkmargin}[1]{\marginnote{#1}{BK}{blue}}

\usepackage{mathtools,bm}

\newcommand{\eat}[1]{}

\newcommand{\wordtovec}{\textsc{Word2Vec}}
\newcommand{\nodetovec}{\textsc{Node2Vec}\xspace}
\newcommand{\moltovec}{\textsc{Mol2Vec}}

\newcommand{\transe}{\textsc{TransE}}
\newcommand{\rescal}{\textsc{Rescal}}
\newcommand{\forward}{\textsc{FoRWaRD}\xspace}
\def\forwardek{$\textsc{FoRWaRD}$\xspace}

\def\forwardekshort{$\textsc{FWD}$\xspace}

\renewcommand{\arraystretch}{1.2}

\def\scs{\sigma}
\def\e#1{\emph{#1}}
\def\key{\mathsf{key}}
\def\dom{\mathsf{dom}}
\def\adom{\mathsf{adom}}

\def\set#1{\mathord{\{#1\}}}

\def\attseq#1{\mathord{\boldsymbol{#1}}}

\def\W{\mathcal{W}}
\def\D{\mathcal{D}}
\def\prob{\mathrm{Pr}}

\def\rel#1{\textit{\emph{\textsc{#1}}}}
\def\att#1{\textit{\emph{\textsf{#1}}}}
\def\val#1{\textit{\emph{\texttt{#1}}}}
\def\qed{\hfill$\blacksquare$}

\newcommand{\authorx}[1]{}
\newcommand{\emailx}[1]{}
\newcommand{\affiliationx}[1]{}

\def\BibTeX{{\rm B\kern-.05em{\sc i\kern-.025em b}\kern-.08em
    T\kern-.1667em\lower.7ex\hbox{E}\kern-.125emX}}
    
\begin{document}

\title{
Stable Tuple Embeddings for Dynamic Databases\\
}

\author{\IEEEauthorblockN{1\textsuperscript{st} Jan Toenshoff}
\IEEEauthorblockA{\textit{RWTH Aachen University} \\
Aachen, Germany\\
toenshoff@informatik.rwth-aachen.de}
\and
\IEEEauthorblockN{2\textsuperscript{nd} Neta Friedman}
\IEEEauthorblockA{\textit{Technion} \\
Haifa, Israel \\
netafr@cs.technion.ac.il}
\and
\IEEEauthorblockN{3\textsuperscript{rd} Martin Grohe}
\IEEEauthorblockA{\textit{RWTH Aachen University} \\
Aachen, Germany\\
grohe@informatik.rwth-aachen.de}
\and
\IEEEauthorblockN{4\textsuperscript{th} Benny Kimelfeld}
\IEEEauthorblockA{\textit{Technion} \\
Haifa, Israel \\
bennyk@cs.technion.ac.il}
}

\maketitle

\begin{abstract}
    We study the problem of computing an embedding of the tuples of a relational database in a manner that is extensible to dynamic changes of the database. 
    In this problem, the embedding should be stable in the sense that it should not change on the existing tuples
    due to the embedding of newly inserted tuples (as database applications might already rely on existing embeddings); at the same time, the embedding of all tuples, old and new, should retain high quality. 
    This task is challenging since inter-dependencies among the embeddings of different entities are inherent in
    state-of-the-art embedding techniques for structured data.
    
    We study two approaches to solving the problem. 
    The first is an adaptation of Node2Vec to dynamic databases. 
    The second is the FoRWaRD algorithm (Foreign Key Random Walk Embeddings for Relational Databases) that draws from embedding techniques for general graphs and knowledge graphs, and is inherently utilizing the schema and its key and foreign-key constraints.
    We evaluate the embedding algorithms using a collection of downstream tasks of column prediction over geographical and biological domains. We find that in the traditional static setting, our two embedding methods achieve comparable results that are compatible with the state-of-the-art for the specific applications. 
    In the dynamic setting, we find that\mgmargin{Just one version, I changed it} the FoRWaRD algorithm generally outperforms and runs faster than the alternatives, and moreover, it features only a mild reduction of quality even when the database
    consists of more than half newly inserted tuples  after the initial training of the embedding.
\end{abstract}


\def\e#1{\emph{#1}}
\section{Introduction}
\begin{figure}
  \centering
  \input{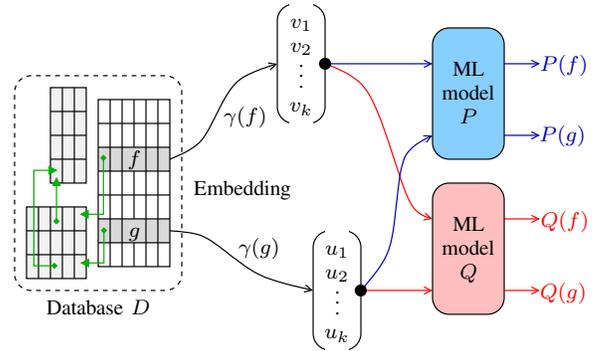}
  \caption{Embedding tuples in a database. Different machine-learning models for various downstream tasks operate over the same vectors that the tuples are mapped to by the embedding. The embedding algorithm can utilize the database structure, particularly the foreign-key references.}\label{fig:dbembed}
\end{figure}

\mgmargindone{I think the fact that we need embeddings does not have to be explained in detail, therefor I have replaced the first paragraph by a single sentence.}
Standard machine learning algorithms assume representations of their input data as numerical vectors. Applying these algorithms for the analysis of non-numerical data requires \emph{embeddings} of these data into a (typically) finite dimensional Euclidean vector space. The embedding needs to be “faithful” to the semantics. 
In particular, similar entities should be mapped to vectors that are close geometrically, and vice versa.
In some modalities, the input comes with a useful embedding to begin with; for example, an image can be represented by the RGB intensities of its pixels. In others, semantic-aware embeddings have to be devised, and indeed have been devised, such as  \wordtovec~\cite{miksutche+13} and RoBERTa~\cite{liu2019roberta} for natural language~\cite{DBLP:conf/icml/LeM14}, \nodetovec~\cite{10.1145/2939672.2939754} and GraphSAGE~\cite{NIPS2017_5dd9db5e} for the nodes of a graph, \transe~\cite{borusugar+13} and RotatE~\cite{sun2019rotate} for the entities of a knowledge graph, and \moltovec~\cite{DBLP:journals/jcisd/JaegerFT18} for molecule structures. 
An approach to evaluating the \e{quality} of a generic embedding technique is via different downstream tasks: solve a collection of machine-learning tasks by utilizing machine-learning models that operate over the embedding (see, e.g.,~\cite{Li2018,DBLP:conf/rep4nlp/LauB16}), as illustrated in Figure~\ref{fig:dbembed}.

Generalizing graph embeddings, generic embeddings have also been devised for relational databases. Such embeddings have enabled the deployment of machine-learning architectures to traditional database tasks such as record similarity~\cite{DBLP:journals/corr/abs-1712-07199, 10.1145/3076246.3076251, DBLP:conf/cidr/BordawekarS19, DBLP:conf/sigmod/Gunther18, DBLP:conf/edbt/0002TNL20}, record linking~\cite{DBLP:conf/sigmod/MudgalLRDPKDAR18,DBLP:journals/pvldb/EbraheemTJOT18} and other integration tasks such as schema, token and record matching (entity resolution)~\cite{DBLP:conf/sigmod/CappuzzoPT20}.  
The embedded entities are typically either tuples or attribute values. 
In this work, we focus on tuple embeddings.
%
\bkmargin{I'm putting Martin's proposal one paragraph later.}
Various approaches have been proposed for obtaining embeddings in databases.  One is to concatenate predefined embeddings of the attribute values (which are, e.g., words or quantities)~\cite{DBLP:conf/sigmod/MudgalLRDPKDAR18,DBLP:journals/pvldb/EbraheemTJOT18} or permutations over the list of values~\cite{DBLP:conf/sigmod/CappuzzoPT20}. Another approach views the tuples as text documents and applies word and document embeddings~\cite{10.1145/3076246.3076251}. Cappuzzo, Papotti and Thirumuruganathan~\cite{DBLP:conf/sigmod/CappuzzoPT20} studied the approach of transforming the database into a graph and applying a node embedding over this graph.  

Arguably, an important advantage of the graph approach to database embedding is that it utilizes information that should be highly relevant to the semantics of the data and is freely available in databases: the \e{structure} of the data. This structure includes more than just the tabular form. An important difference between embedding a single relation and embedding a full relational database is that the latter entails considerable semantics via dependencies between relations: typically, columns in a database store foreign keys, which have no semantic meaning as atomic values but may have rich semantic meaning as references to tuples from other relations. 
For instance, we can infer considerable information about a tuple with apparently meaningless values, such as internal codes, by looking at the tuples that are referenced by this tuple. 
Hence, to appropriately capture the semantics of the data, embedding methods need to incorporate such dependencies.
Yet, we are not aware of any embedding technique that directly uses the most common way of referencing tuples, namely foreign-key references. Yet, it has been shown useful in the general context of machine learning~\cite{DBLP:conf/cidm/SchulteBCBX13,shah2017key}. As we explain later on, we show here how these can be gracefully utilized for the sake of high-quality tuple embedding.

Incorporating the database structure, and particularly references among tuples, means that the embeddings of tuples depend upon each other. 
This leads to new challenges since the database is often not a static object but rather serves a dynamic organization (e.g., with arrival of new customers and purchases, new patients and patient records, etc.). When new data arrives, we are in a situation where we have an embedding for the old tuples but not the new ones. A straightforward solution to this problem is to reapply the embedding algorithm from scratch over the new database. This approach, however, suffers from two main drawbacks. First, it might be computationally too expensive to compute the embedding over the entire database upon every tuple arrival. More fundamentally, reapplying the embedding algorithm  is likely to change the embedding of the old tuples due to the inherent randomness in most embedding algorithms.

It has been empirically demonstrated~\cite{schumwolrit+21,wang2020instability} that for many standard node embedding algorithms for graphs, the embeddings change considerably even when applied repeatedly to identical graphs with identical parameter settings. Even worse, standard downstream classifiers change their predictions for individual data points when applied to these embeddings~\cite{schumwolrit+21}. A similar behaviour has been observed earlier for word embeddings \cite{hellrich2016bad,antoniak2018stability,leszczynski_understanding_2020}. Clearly, such an instability, in particular when it even changes downstream classification results, is hard to tolerate, even more so if the database changes often and hence the embedding needs to be updated frequently. However, we cannot just ignore changes in the database, and we need to insert new tuples into the embedding as they arrive to keep the downstream tasks up to date.

The most pragmatic and feasible solution to this dilemma is to freeze the embedding of the existing tuples in the database and only compute embeddings for new tuples as they arrive dynamically. Hence, we address the \e{stable} variant of the problem where the goal is to infer embeddings of new tuples \e{without changing the embedding of old ones}. Of course, the challenge is to do so while retaining high quality of the embedding. One may suspect that the performance of downstream tasks will suffer significantly when based on a stable embedding of a dynamically changing database. Yet we demonstrate that with the right embedding algorithms, this is not the case.


To be more precise, we study the following task that we refer to as the \e{stable database embedding} problem.
We need to devise two algorithms. The first algorithm, applied in the \e{static phase}, takes as input a database $D$ over a schema $\scs$ and learns a tuple embedding $\gamma$ that maps every fact of $D$ (i.e., occurrence of a tuple in a certain relation in $D$) to the vector space $\mathbb R^k$ (for some hyperparameter $k$).
Note that this phase solves the task of \e{static database embedding}, that is, the embedding problem in its traditional sense.
The second algorithm, applied in the \e{dynamic phase}, has access to $D$ and the tuple embedding $\gamma$ and takes as input a newly arrived tuple $t$ that is not in $D$; the goal of this algorithm is to extend $\gamma$ to $D\cup\set{t}$ by determining the value $\gamma(t)$.  Importantly, the schema $\scs$ specifies the key and foreign-key constraints that can be used for understanding the actual foreign-key references that exist inside the database $D$ and its future evolution.

Since we wish to keep the embedding of the existing tuples stable (for the aforementioned reasons), deleting tuples from a database is not an issue---we simply delete the tuples and their images under the embedding. This is why we focus on tuple insertion in our framework and experiments.  Of course, eventually there may be a point where the database has changed so much that a completely new embedding has to be computed.

\subsubsection*{Contribution}
We design two main solutions to the stable database embedding problem. The first is an adaptation of \nodetovec, and the second is what we call the \forward algorithm.

\paragraph{\nodetovec adaptation.} 
The first solution that we design is based on our adaption of \nodetovec to relational database embeddings.
We found empirically that this adaptation performs very well for the static embedding problem.
For the stable database embedding problem, we devise a dynamic version of \nodetovec that is based on the following idea. When we extend an existing embedding to new nodes, we sample relevant paths and continue the training of \nodetovec from where it stopped while performing gradient descent \e{only on the embeddings of new nodes}.

\paragraph{The \forward algorithm.}
The second solution we devise is a new algorithm, \forward (\underline{Fo}\-reign Key \underline Random \underline{Wa}lk Embeddings for \underline Relational \underline Databases), that is inherently built to accommodate the structure of relational databases and, importantly, to be extensible to dynamic databases.  We demonstrate experimentally that it performs very well for the stable database embedding problem, which it was designed for, as well as the static database embedding problem.
\forward draws from node embedding techniques based on random walks \cite{10.1145/2939672.2939754,peralrski14} as well as the knowledge graph embedding algorithm \rescal~\cite{nictrekri11}. From each tuple in the database we start random walks in the database by repeatedly following foreign-key references. We embed tuples of the database depending on the similarity of the distributions of these walks, where we measure similarity in terms of a predefined similarity measures on the attribute values. 
\bkmargin{Rephrasing the following}
Crucially, while learning the embedding of tuples, we also learn a separate similarity measure for each type (relational scheme) of a walk. This similarity is encoded as a matrix describing an inner product.
Once we have learned an initial static embedding, we can dynamically extend it to each new tuple by essentially solving a system of linear equations that constrain the distances to the existing tuples (or a random subset of these).

We describe our
experiments over several downstream tasks of (binary and multi-label) column prediction from multiple benchmark databases over geographical and biological domains~\cite{DBLP:conf/kdd/NevilleJFH03, 10.5555/2073876.2073934, doi:10.1021/jm00106a046, Lodhi05ismutagenesis}. We show that \forward performs very well even for the static database embedding problem. 
On the majority of the benchmarks, both \forward and our static adaptation of \nodetovec clearly outperform the state-of-the-art baselines (see Table~\ref{table:static}).

Regarding the stable database embedding problem, both dynamic \nodetovec and \forward perform well, though overall \forward shows a superior performance (as we show in Figure~\ref{fig:dynamic_results} and Table~\ref{table:dynamic}).  It is also faster than \nodetovec. Quite remarkably, the performance of \forward turns out to be rather stable as we add more and more tuples to the database. Even with 50\%, sometimes even 80\%, newly added tuples the drop in accuracy for column prediction is small (see Figure~\ref{fig:dynamic_results}). Combined with the strong performance of \forward\ for the static embedding problem, this convincingly shows that \forward is a viable solution to the stable embedding problem.

In summary, our contributions are as follows. First, we define the problem of stable database embedding.  Second, we devise an adaptation of \nodetovec to dynamic databases. Third, we devise the \forward algorithm in both its static and dynamic versions.  Fourth, we conduct a thorough experimental evaluation of our algorithms over a collection of tuple prediction benchmarks.

\subsubsection*{Organization} The remainder of the paper is organized as follows. After giving preliminary definitions and notation in Section~\ref{sec:preliminaries}, we define the problem of stable database embedding in Section~\ref{sec:problem_definition}. 
In Section~\ref{sec:node2vec} we describe our \nodetovec adaption, and in Section~\ref{sec:method} the \forward algorithm.
Finally, we present our experimental evaluation in Section~\ref{sec:experiments} and conclude in Section~\ref{sec:conclusions}.


\section{Preliminaries}\label{sec:preliminaries}
We focus on databases over schemas with key and foreign-key constraints. More precisely, a \e{database schema} $\scs$ consists of a finite collection of \e{relation schemas} $R(A_1,\dots,A_k)$ where $R$ is a distinct \e{relation name} and each $A_i$ is a distinct \e{attribute name}. For simplicity, we assume that the attribute sets of distinct relations are disjoint.
Each attribute $A$ is associated with a  \e{domain}, denoted $\dom(A)$. Each relation schema $R(A_1,\dots,A_k)$ has a unique \e{key}, denoted $\key(R)$, such that $\key(R)\subseteq\set{A_1,\dots,A_k}$. 

A \e{foreign-key constraint} (FK) is  an inclusion dependency of the form $R[\attseq B]\subseteq S[\attseq C]$ where $R$ and $S$ are relation names, $\attseq B=B_1,\dots, B_\ell$ and $\attseq C=C_1,\dots,C_\ell$ are sequences of distinct attributes of $R$ and $S$, respectively, and $\key(S)=\set{C_1,\dots,C_\ell}$.

A \e{database} $D$ over the schema $\scs$ is a finite set of \e{facts} $R(a_1,\dots,a_k)$ over the relation schemas $R(A_1,\dots,A_k)$ of $\scs$, so that $a_i\in\dom(A_i)$ for all $i=1,\dots,k$. In addition, such an $a_i$ can be missing, in which case we assume that it is a distinguished \e{null} value (that belongs to none of the attributes domains) denoted by $\bot$.  The fact $R(a_1,\dots,a_k)$ is also called an \e{$R$-fact} and a \e{$\scs$-fact}. We denote by $R(D)$ the restriction of $D$ to its $R$-facts.  For a fact $f=R(a_1,\dots,a_k)$ over $R(A_1,\dots,A_k)$, we denote by $f[A_i]$ the value $a_i$, and by $f[B_1,\dots,B_\ell]$ the tuple $(f[B_1],\dots,f[B_\ell])$. For a database $D$, the \e{active domain} of an attribute $A$ (\e{w.r.t.~to $D$}), denoted $\adom_D(A)$ or just $\adom(A)$ if $D$ is clear from the context, is the set of values that occur in $D$ for the attribute $A$, that is, $\adom(A)=\set{f[A]\mid f\in R(D)}$.

We require that the database $D$ over $\scs$ satisfies the constraints of $\scs$. In particular, for the key constraints we require every two distinct $R$-facts $f_1$ and $f_2$ must satisfy $f_1[C]\neq f_2[C]$ for at least one attribute $C\in\key(R)$, and both  $f_1[C]$ and $f_2[C]$ are nonnull. Moreover, for every FK $\varphi$ of the form $R[\attseq B]\subseteq S[\attseq C]$ and $R$-fact $f\in D$, if $f[\attseq B]$ has no nulls\footnote{Note that we adopt the convention that an FK is ignored in a fact that includes nulls in one or more of the referencing attributes.} then there exists an $S$-fact $g\in D$ such that $f[\attseq B]=g[\attseq C]$; in this case, we say that \e{$f$ references $g$ via $\varphi$}
(note that $f$ references precisely one fact via $\varphi$). 

\begin{example}
Figure~\ref{fig:dbexample} depicts an example of a movie database over a schema. 
The leftmost column of each relation includes tuple names for later reference, and is not considered part of the database itself.
As conventional, keys are marked by underlining the key attributes. The FKs of each relation schema are given under the corresponding relation. For example, 
The \rel{Movies} relation has the key constraint
$\key(\rel{Movies})=\set{\att{mid}}$
and the FK $\rel{Movies}[\att{studio}]\subseteq\rel{Studio}[\att{sid}]$. The reader can verify that the FK is indeed satisfied by the database of the figure; for example,
\val{s03} is indeed the sid attribute of a fact of 
$\rel{Studio}$, namely $s_3$.
Finally, observe that the genre attribute of $m_3$ is missing, that is, $m_3[\att{genre}]=\bot$.
\qed
\end{example}




\section{Problem Definition}\label{sec:problem_definition}

\newcommand{\cc}{\cellcolor{lightgray}}

\begin{figure*}[t]
  \centering
  \parbox{3.5in}{\centering
  \hfill\begin{tabular}[t]{l|c|c|c|c|c|}
         \multicolumn{1}{c}{} & \multicolumn{5}{l}{\rel{Movies}}\\  \cline{2-6}
         &\cc \underline{\att{mid}} &\cc \att{studio} &\cc
                                                               \att{title} &
                                                                       \cc \att{genre} & \cc \att{budget} \\
        \cline{2-6}
        $m_1$ & \val{m01} & \val{s03} & \val{Titanic} & \val{Drama} & \val{200M }\\
        $m_2$ & \val{m02} & \val{s01} & \val{Inception} & \val{SciFi} & \val{160M} \\
        $m_3$ & \val{m03} & \val{s01} & \val{Godzilla} & $\bot$  & \val{150M} \\
        $m_4$ &\val{m04} & \val{s03} & \val{Interstellar} & \val{SciFi} & \val{160M} \\
        $m_5$ & \val{m05} & \val{s02} & \val{Tropic Thunder} & \val{Action} & \val{90M} \\
       $m_6$ &  \val{m06} & \val{s01} & \val{Wolf of Wall St.} & \val{Bio} & \val{100M} \\
       \hline
       \multicolumn{6}{|l}{}\\
       \multicolumn{6}{|l}{$\rel{Movies}[\att{studio}]\subseteq \rel{Studios}[\att{sid}]$}\\\hline
  \end{tabular}
  \vskip2em
  \hfill\begin{tabular}[t]{l|c|c|c|}
        \multicolumn{1}{c}{} & \multicolumn{3}{l}{\rel{Actors}}\\
        \cline{2-4}
        & \cc \underline{\att{aid}} & \cc \att{name} & \cc \att{worth} \\
        \cline{2-4}
        $a_1$ & \val{a01} & \val{DiCaprio} & \val{230M} \\
        $a_2$ & \val{a02} & \val{Watanabe} & \val{40M} \\
        $a_3$ & \val{a03} & \val{Cruise} & \val{600M} \\
        $a_4$ & \val{a04} & \val{McConaughey} & \val{140M} \\
        $a_5$ &  \val{a05} & \val{Damon} & \val{170M} \\
        \hline
    \end{tabular}
    }\hskip5em
  \parbox{2.5in}{
    \begin{tabular}[t]{l|c|c|c|}
        \multicolumn{1}{c}{} &   \multicolumn{3}{l}{\rel{Studios}}\\ \cline{2-4}
        &\cc \underline{\att{sid}} & \cc \att{name} & \cc \att{loc} \\
         \cline{2-4}
        $s_1$ & \val{s01} & \val{Warner Bros.} & \val{LA} \\
        $s_2$ & \val{s02} & \val{Universal} &\val{ LA} \\
        $s_3$ &  \val{s03} & \val{Paramount} & \val{LA} \\
       \hline
    \end{tabular}
    \vskip4em
    \begin{tabular}[t]{l|c|c|c|l}
        \multicolumn{1}{c}{} &\multicolumn{3}{l}{\rel{Collaborations}}\\
        \cline{2-4}
         & \cc \underline{actor1} & \cc \underline{actor2} & \cc \underline{movie} \\
       \cline{2-4}
        $c_1$ & \val{a01} & \val{a02} & \val{m03} \\
        $c_2$ & \val{a04} & \val{a05} & \val{m04} \\
        $c_3$ & \val{a04} & \val{a03} & \val{m05} \\
        $c_4$ & \val{a01} & \val{a04} & \val{m06} \\
        \cline{1-4}
        \multicolumn{5}{|l}{}\\
       \multicolumn{5}{|l}{$\rel{Collaborations}[\att{actor1}]\subseteq\rel{Actors}[\att{aid}]$}\\
       \multicolumn{5}{|l}{$\rel{Collaborations}[\att{actor2}]\subseteq\rel{Actors}[\att{aid}]$}\\
       \multicolumn{5}{|l}{$\rel{Collaborations}[\att{movie}]\subseteq\rel{Movies}[\att{mid}]$}\\
       \hline
    \end{tabular}
    }
    \caption{\label{fig:dbexample}Database example. Key constraints are marked with underline and foreign-key references are specified under the corresponding relations.}
\end{figure*}

As explained in the Introduction, the problem of stable database embedding consists of two tasks.
\begin{itemize}
\item \e{Static phase}: The goal is to derive an embedding of the tuples in the traditional sense. Formally, we are given a database $D$ over a schema $\scs$, and we wish to compute an embedding function $\gamma: D\rightarrow \mathbb R^k$ for some hyperparameter $k>0$. 
\item \e{Dynamic phase}: Here the goal is to extend the embedding $\gamma$ to a new fact $f$. Formally, we are given a database $D$ over a schema $\scs$, a precomputed embedding $\gamma: D\rightarrow \mathbb R^k$ and a new fact $f\notin D$. Our goal is to compute a new embedding $\gamma': D\cup\set{f}\rightarrow \mathbb R^k$ such that $\gamma'(f')=\gamma(f')$ for all $f'\in D$. Hence, we only need to compute $\gamma'(f)$.
\mgmargin{$f'\to f$}
 \end{itemize}
 In the obvious manner, we can generalize the dynamic phase to a set (\e{batch}) $\{f_1,\dots,f_\ell\}$ of new facts rather than just a single $f$, and the goal is to extend the embedding to all of the facts of this set. \mgmargin{Maybe we should remark that sometimes it is necessary to insert in batches to keep the constraints satisfied.}
 
 The objective is to compute an embedding $\gamma$ that represents the data in a way that makes it accessible for data analysis and machine learning algorithms. To evaluate an embedding algorithm, we empirically test it against various downstream learning tasks on the relational data. That is, we take the embedded data as input for a machine learning algorithm (e.g., an artificial neural network) and measure how well it performs. To evaluate the performance of the dynamic embedding algorithm, we measure the performance not only over the original database, on which the machine learning algorithm was trained, but also on the tuples newly added as the database changes dynamically. 
  
We need to devise two algorithms: one for the static phase and one for the dynamic phase. Note that these algorithms might (and actually should) depend on each other: the algorithm for the static phase not only needs to perform well as a static embedding algorithm, but also allow and enhance the effectiveness of the algorithm for the dynamic phase.
 
\begin{example}
Let $D'$ be the database of Figure~\ref{fig:dbexample} and $D=D'\setminus{c_4}$. Hence, $D'$ is obtained from $D$ by inserting the fact
\[c_4=\rel{Collaborations}(\val{a01},\val{a04},\val{m06})\,.\]
In static phase, for the input $D$ we need to compute a mapping $\gamma:D\rightarrow\mathbb R^k$. In the dynamic phase, when inserting $c_4$ into $D$ we wish to extend $\gamma$ to $c_4$ by computing $\gamma(c_4)$ without changing it on the facts of $D$; for example, $\gamma(c_1)$ and $\gamma(m_1)$ should remain intact. To determine $\gamma(c_4)$ we can utilize the semantic knowledge that the new $c_4$ references the existing $a_1$, $a_4$ and $m_6$.
\qed
\end{example}


\section{Dynamic Database Embeddings Based on Node2Vec}\label{sec:node2vec}
The first algorithm we propose for the stable database embedding problem is an adaptation of \nodetovec, a well-known algorithm for node embeddings of graphs \cite{10.1145/2939672.2939754}. In addition to the original \nodetovec, we incorporate ideas from \cite{DBLP:conf/sigmod/CappuzzoPT20}, where \nodetovec\ was adapted to embedding static relational databases, and ideas from \cite{mahavi2018dynnode2vec}, where \nodetovec\ was adapted for node embeddings of dynamic graphs. A crucial novelty in our approach is that we incorporate foreign key constraints by identifying certain nodes in the graph model of the relational database to which we apply \nodetovec.

\begin{figure}
  \newcommand{\angs}[1]{$\langle#1\rangle$}

    \centering
    \input{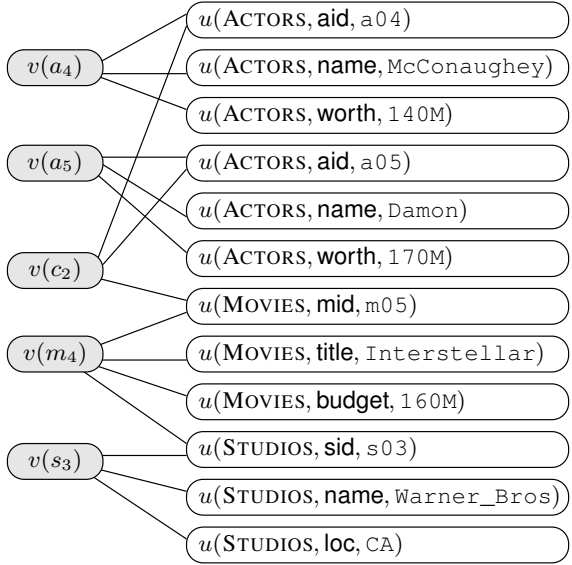}
    \caption{A partial graph of the database in Figure~\ref{fig:dbexample}.}
    \label{fig:n2vgraph}
\end{figure}

We start with a description of the static embedding algorithm. For the rest of the section, let $D$ be a database of schema $\sigma$.
We build a bipartite graph $G_D$ such that one side represents the facts and the other side represents the attribute values that occur in the facts, as illustrated in
Figure~\ref{fig:n2vgraph} for a fragment of the database of Figure~\ref{fig:dbexample}. The graph is related to, but slightly different from the graph in 
\cite{DBLP:conf/sigmod/CappuzzoPT20}. 
For each relation schema $R(A_1,\ldots,A_k)$ in $\sigma$, each attribute $A_i$, and each value $a$ of $A_i$ that occurs in $R(D)$, we add a node $u(R,A_i,a)$ to $G_D$. For each fact $f=R(a_1,\ldots,a_k)$ in $R(D)$ we add a node $v(f)$ and, for $i=1,\ldots,k$ we add edges between $v(f)$ and  
$u(R,A_i,a_i)$. So far, the components of the graph for different relations $R,S$ in $\sigma$ are disconnected and completely independent. In a second step, we use the foreign key constraints to connect these subgraphs. Suppose we have an FK $R[B_1,\ldots,B_\ell]\subseteq S[C_1,\ldots,C_\ell]$. Then for all $i\in[\ell]$ and all values $a\neq\bot$, we identify the two nodes $u(R,B_i,a)$ and $u(S,C_i,a)$ if they both exist.
Figure~\ref{fig:n2vgraph} illustrates the construction.

Note that we are careful in the way we introduce connections in our graph: if the same value occurs in different columns or different relations, there is no connection between the two occurrences, except if they are linked by a foreign key. To understand why this is the correct way of modeling the database by a graph, just imagine that in our movie database of Figure~\ref{fig:dbexample} we had a tuple $m_7$ representing a movie called "Universal". Then the occurrence of \texttt{Universal} in the \textsf{title} column of the \textsc{Movies} relation would be completely independent of the occurrence of \texttt{Universal} in the \textsf{name} column of the \textsc{Studios} relation. So it is appropriate to represent these two by distinct nodes in the graph. However, the occurrence of \texttt{s01} in the \textsf{studio} column of the \textsc{Movies} relation is referring to the same object as the occurrence of \texttt{s01} in the \textsf{sid} column of the \textsc{Studios} relation, and therefore it is justified to identify the two nodes.

\subsection{Extension to the Dynamic Setting}\label{sec:n2vdynamic}
We now describe our extension of \nodetovec\ to the dynamic setting. Beres et al.~\cite{beres2019node} suggested an online extension of \nodetovec\ on graphs in the setting where the graph arrives as a stream of edges, and Mahdavi et al.~\cite{mahavi2018dynnode2vec} suggested a method based on evolving walks generation, in the setting where the input is a series of graphs that arrive at discrete timestamps. Neither of these applies directly to the stable embedding problem, because is does not allow to freeze of the old embeddings
(i.e., in these methods the old embeddings change when new nodes arrive). Nevertheless, we use similar ideas as \cite{mahavi2018dynnode2vec}.

Our method is based on sampling new walks and continuing the gradient descent. Suppose we have already trained a node embedding for the graph $G_D$ associated with our current database $D$. As a new fact $f$ arrives, we update the graph to $G_{D'}$ for the new database $D\cup\{f\}$. Note that the new nodes in $V(G_{D'})\setminus V(G_D)$ are $v(f)$ and possibly so nodes $u(R,A,a)$ for values $a$ in $f$ that have not been present before. We sample new random walks starting at the new nodes. Then we train a new \nodetovec\ model for $G_{D'}$ taking the old model for $G_D$ and a random initialization for the new nodes as the intialization. As we train the new model by standard gradient descent techniques, we freeze the old nodes and only update the embedding on the new nodes.

\section{The \forward Algorithm}\label{sec:method}
We now present our second algorithm for the stable database embedding problem. 
Our embedding algorithm will use \emph{random walks} along foreign-key constraints to incorporate the structure of the database into the embedding, and it will use \emph{kernelized domains} to incorporate the semantics of the values appearing in the facts. 
Before we describe the algorithm, we establish the background on random walks and on kernelized domains in Sections~\ref{sec:random-walks} and~\ref{sec:kernels}.

Throughout this section, we fix a database schema $\sigma$, and we always assume that $D$ is a database of schema $\sigma$. 

\subsection{Random Walks over Database Facts}
\label{sec:random-walks}
We consider random walks over database facts, where the transition
from one fact to another follows a pattern (or scheme) of FK (forward
or backward) references. Formally, a \emph{walk scheme}
is a sequence $s$ of the form
\begin{eqnarray}
\label{eq:walkscheme}
R_0[\attseq A^0]\mbox{---}R_1[\attseq B^1]\,,\,\!
R_1[\attseq A^1]\mbox{---}R_2[\attseq B^2]\,,\,
\dots \\
,R_{\ell-1}[\attseq A^{\ell-1}]\mbox{---}R_{\ell}[\attseq B^{\ell}] \notag
\end{eqnarray}
such that for all $k=1,\dots,\ell$, either
$R_{k-1}[\attseq A^{k-1}] \subseteq R_{k}[\attseq B^k]$ is an FK or
$R_{k}[\attseq B^k] \subseteq R_{k-1}[\attseq A^{k-1}]$
is an FK.
We say that $s$ has \e{length} $\ell$, that it \e{starts from} $R_0$ and that it \e{ends with} $R_\ell$.

\begin{figure*}
  \input{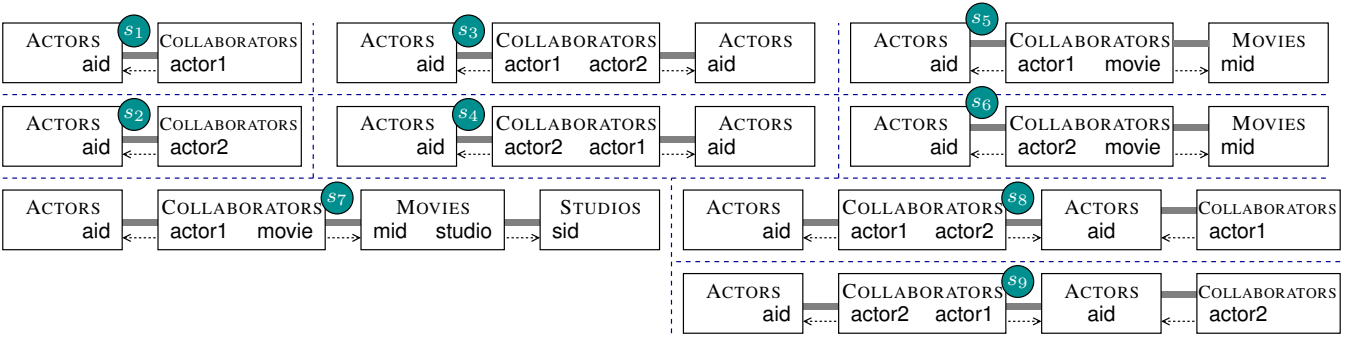}
  \caption{\label{fig:walks} All walk schemes of length at most three, for the database schema of Figure~\ref{fig:dbexample}, that start from the Actor relation.}
\end{figure*}

\begin{example}
Figure~\ref{fig:walks} depicts nine walk schemes, $s_1,\dots,s_9$ that
start from the \rel{Actor} relation. The reader can verify that these are
all of the walk schemes of length at most three that start from
\rel{Actor}. 
For illustration, let us consider the scheme $s_5$ from the figure. In our notation, this walk scheme is written as
\begin{align*}
\rel{Actors}[\att{aid}]&\mbox{---}\rel{Collaborations}[\att{actor2}], \\
\rel{Collaborations}[\att{movie}]&\mbox{---}\rel{Movies}[\att{mid}]\,.
\end{align*}
Note that $s_1$ ends with \rel{Collaborations} while $s_5$ ends with with \rel{Movies}. 
\qed
\end{example}

A \emph{walk} with the scheme $s$ is a sequence $(f_0,\dots,f_{\ell})$
of facts such that $f_k$ is an $R_k$-fact and
$f_{k-1}[\attseq A^{k-1}]=f_{k}[\attseq B^{k}]$ for all $k=1,\dots,\ell$.  
We say that
$(f_0,\dots,f_\ell)$ \e{starts from}, or has the \e{source},
$f_0$, and that it \e{ends with}, or has the \e{destination},
$f_\ell$.

\begin{example}\label{example:walks}
Continuing our example, consider again the walk scheme $s_5$
of Figure~\ref{fig:walks}.
Starting at the fact $a_1$ of Figure~\ref{fig:dbexample}, there are two walks that follow the scheme $s_5$, namely $(a_1,c_1,m_3)$ and $(a_1,c_4,m_6)$.
\qed
\end{example}

Note that we allow walk schemes and walks of length zero.
For each relation $R$ there is a scheme of length zero that starts and ends in $R$.
The walks of this scheme have the form $(f_0)$ and simply end directly at the start fact $f_0$ in $R$.

Let $s$ be a walk scheme as written in~\eqref{eq:walkscheme}. 
By a \emph{random walk} with the scheme $s$ we refer to the walk obtained
by uniformly selecting the next valid fact in the walk. 
More formally, let $f_0=f$ be an $R_0$-fact. We denote by $\W(f,s)$ 
the distribution over the walks with the walk scheme $s$ where each walk is sampled by
starting from $f_0$ and then iteratively selecting $f_{k}$, for
$k=1,\dots,\ell$, randomly and uniformly from the set
$\{f\in R_{k}\mid f[\attseq B^k]=f_{k-1}[\attseq A^{k-1}]\}$.
We denote by $d_{f,s}$ the random variable/element that maps each walk in $\W(f,s)$ to its destination, that is, the last fact in the walk. Then for a fact $g\in R^k(D)$, the probability that a walk sampled from $\W(f,s)$ ends with $g$ is $\prob(d_{f,s}=g)$.
Observe that for every attribute $A$ of $R_k$ we get the random variable $d_{f,s}[A]$ that forms the value of the random walk's destination in the attribute $A$.
Given a start fact $f_0$ in the start relation $R_0$ of walk scheme $s$, one can compute the distribution $d_{f,s}$ through a simple breadth first search along the sequence of foreign keys specified by $s$. 

\begin{example}
Recall from Example~\ref{example:walks} that the walks
$(a_1,c_1,m_3)$ and $(a_1,c_4,m_6)$ are the only two walks that follow the scheme $s_5$ (Figure~\ref{fig:walks})
and start from $a_1$ (Figure~\ref{fig:dbexample}).
Therefore, for the random variable $d_{a_1,s_5}$ it holds that
$\prob(d_{a_1,s_5}=m_3)=0.5$ and $\prob(d_{a_1,s_5}=m_6)=0.5$.
Moreover, we have the following.
\begin{align*}
    \prob(d_{a_1,s_5}[\att{budget}]&=150\text{M})=0.5\\
    \prob(d_{a_1,s_5}[\att{budget}]&=100\text{M})=0.5\\
    \prob(d_{a_1,s_5}[\att{genre}]&=\text{Bio})=1.0
\end{align*}
Hence, each of  $d_{a_1,s_5}[\att{budget}]$ and
$d_{a_1,s_5}[\att{genre}]$ indeed defines a distribution over attribute values.
\qed
\end{example}

Recall that databases (in real life and in our formal framework) may have missing values.
A random walk starting at $f$ might end at a fact $R_\ell$-fact $g$ which has no known value for an attribute $A$ of $R_\ell$.
Therefore, the random variable $d_{f,s}[A]$ can in theory assume the value $\bot \not\in dom(A)$.
As a convention, we define the probability distribution of $d_{f,s}[A]$ as the posterior distribution after $d_{f,s}[A] \neq \bot$ is known.
With this modification we enforce $d_{f,s}[A] \in dom(A)$.
This will be crucial in Section \ref{sec:kernels}, where we define similarity measures for $d_{f,s}[A]$ based on $dom(A)$.
If all walks from $f$ with scheme $s$ end at facts $g$ with $g[A]=\bot$, 
then $d_{f,s}[A]$ does not exist and is not considered by \forward{}.
This also includes the case where no walks with scheme $s$ exist from start $f$.\mgmargin{Maybe we should say this earlier.}

\subsection{Kernelized Domains}
\label{sec:kernels}
In addition to the structural information carried by the foreign-key random walks described in the previous section, our embeddings are based on similarities between values occurring in the tuples. Formally, we assume that these similarities are given by \emph{kernels}. For every attribute $A$ occurring in the database schema $\sigma$ we assume that we have a symmetric binary function $\kappa_{A}$ mapping pairs of elements from $\dom(A)$ to the nonnegative reals. Intuitively, $\kappa_{A}(a,b)$ measures the similarity between elements $a,b\in\dom(A)$. Formally, $\kappa_{A}$ needs to satisfy certain properties that turn it into a kernel function. This means that there is an embedding $\alpha_{A}: \dom(A) \rightarrow \mathcal{H}_{A}$, where $\mathcal{H}_{A}$ is a Hilbert space, that is, a (possibly infinite) vector space with an inner product, which we denote by $\langle\cdot,\cdot\rangle$, and $\kappa_{A}$ is defined by $\kappa_{A}(a,b)=\langle\alpha_{A}(a),\alpha_{A}(b)\rangle$. Importantly, the mapping $\alpha_{A}$ is only implicit, our algorithms only need access to the kernel function $\kappa_{A}$. For details on kernels, we refer the reader to \cite[Chapter~16]{DBLP:books/daglib/0033642}, \cite{hofmann2008kernel}.

For many natural domains, such a kernel function can be obtained by standard embedding techniques. For examples, we can use word embeddings, possibly tailored towards specific applications, for natural language domains \cite{DBLP:conf/icml/LeM14}, or molecule embeddings in the case
of a domain of molecules~\cite{DBLP:journals/jcisd/JaegerFT18}. For numerical domains, we can use a Gaussian kernel
$\kappa(a,b) = \exp\left(-(a-b)^2/2\upsilon\right)$ for some "variance" $\upsilon>0$. As a fallback, we can always use the \emph{equality kernel} defined by $\kappa(a,a)=1$ and $\kappa(a,b)\neq 0$ for $a\neq b$. We would typically use the equality kernel for finite categorical domains and domains consisting of identifiers, which have no semantic meaning.

Kernel functions offer a straightforward way of encoding domain knowledge by modeling the similarity of the domain values.
Kernels are also helpful when dealing with noisy data.
For example, on text, kernels based on the edit distance can be used to smooth out random typos.
In the following we use these kernel functions to define similarity measures for the random variables $d_{f,s}[A]$.

Let $s$ be a walk scheme of length $\ell$ from $R$ to
$R'$. Let $A$ be an attribute of $R'$ and let $f$ and $f'$ be
two distinct $R$-facts. Under our assumption,\mgmargin{Which assumption? And what exactly is the distribution?} $d_{f,s}[A]$ and
$d_{f',s}[A]$ are random variables over a shared kernelized domain
$dom(A)$.  We utilize this to quantify the similarity between
$d_{f,s}[A]$ and $d_{f',s}[A]$ with respect to the underlying kernel
$\kappa_{A}$.  
To this end we define the \emph{Expected Kernel Distance}
\text{KD} is the expected distance between two random values
selected independently at random:
\begin{align}
    \text{KD}(d_{s,f}[A],\!d_{s,f'}[A]) \label{eq:KEK}
    \!&=\! \underset{\W(f,s)\times\W(f',s)}{\mathbb{E}}\hspace{-2em}[\kappa_A(d_{s,f}[A],\!d_{s,f'}[A])]
\end{align}


\subsection{Embedding}
Let us now describe our static embedding algorithm. Recall that $D$ denotes a database of schema $\sigma$. We describe how to embed the tuples of a single relation $R(D)$; of course we can then apply the method to embed all relations in $D$. So our goal is to compute a vector embedding $\varphi: R(D) \mapsto \mathbb{R}^d$ of dimension $d \in \mathbb{N}$.
Intuitively, \forward{} embeddings aim to model the similarity of the random walk destinations  $d_{s,f}$ for all $R$-facts $f$ and all walk schemes $s$ starting from $R$ of length up to a certain walk length $\ell_\text{max}$.
Formally, we define $\mathcal{T}(R, \ell_\text{max})$ as the set of all pairs $(s,A)$ such that $s$ is a walk scheme of length at most $\ell_{\text{max}}$ starting from $R$, and $A$ is an attribute of the destination $R'$ of $s$ that is not involved in any foreign-key constraints. 
Together with $\phi$ we compute an auxiliary embedding $\psi:\mathcal{T}(R, \ell_\text{max}) \rightarrow \mathbb{R}^{d \times d}$ that maps each pair $(s,A)$ to a symmetric matrix $\psi(s,A)$.

Ideally, our goal is to find $\varphi$, $\psi$ satisfying
\begin{equation}
    \label{eq:objective}
    \varphi(f)^{\top}\psi(s,A)\varphi(f') = \text{KD}(d_{s,f}[A], d_{s,f'}[A])
\end{equation}
for all $f,f'\in R(D)$ and $(s,A)\in \mathcal{T}(R, \ell_\text{max})$. 
Usually, we will not be able to achieve this. Instead, we aim at jointly minimising 
\begin{equation}
    \label{eq:objective-min}
    \big|\varphi(f)^{\top}\psi(s,A)\varphi(f') - \text{KD}(d_{s,f}[A], d_{s,f'}[A])\big|
\end{equation}
for all $f,f',s,A$. Then the embedding $\varphi$ is the primary output of \forward.

Essentially, what we do here is learn an inner product $\langle \cdot,\cdot\rangle_{S,A}$ on the latent space of the embedding $\phi$ defined by $\langle x,y\rangle_{S,A}=x^\top \psi(s,A) y$ for all $s$ and $A$, and we try to find a $\phi$ such that the similarity of facts $f$ and $f'$ with respect to this inner product matches the similarity between the random variables $d_{f,s}[A]$ and $d_{f',s}[A]$ with respect to the underlying kernel
$\kappa_{R_\ell.A}$. The idea of learning auxiliary inner products (or their matrices) jointly with the actual embedding goes back to the knowledge graph embedding algorithm \rescal~\cite{nictrekri11}.

It remains to describe the optimisation procedure that we use for minimising \eqref{eq:objective-min}. 

\subsection{Optimization}
\label{sec:opt}
We utilize gradient descent to optimize \forward{} embeddings.
That is, the embeddings $\varphi$ and $\psi$ are initialized randomly and then stochastic gradient descend is used to minimize the $\ell_2$-loss of the objective in Equation \ref{eq:objective-min}.
During training, we sample a large number of tuples of the form $(f, f', s, A, g, g')$.
Here, $f$ and $f'$ are $R$ facts from the database and $(s,A) \in \mathcal{T}(R, \ell_\text{max})$.
The $R_\ell$ facts $g$ and $g'$ are the destinations of random walks with scheme $s$ sampled for $f$ and $f'$, respectively.
To this end, we specify a hyperparameter $n_\text{samples} \in \mathbb{N}$.
For each $R$-fact $f$ and each $(s,A) \in \mathcal{T}(R, \ell_\text{max})$ for which $d_{s,f}[A]$ exists we uniformly sample $n_\text{samples}$ of the form $(f, f', s, A, g, g')$ with $f' \neq f$.
If the total number of such samples is less than $n_\text{samples}$, then we just use all samples without duplicates.
Using these samples, we minimize the following term with stochastic gradient descent:
\begin{equation}
    \mathcal{L} = \frac{1}{2}|\varphi(f)^{\top}\psi(s,A)\varphi(f') - \kappa_{R_\ell.A}(g[A], g'[A])|^2.
\end{equation}
This objective uses the value $\kappa_{R_\ell.A}(g[A], g'[A])$ as a (stochastic) estimate of $\text{KD}(d_{s,f}[A], d_{s,f'}[A])$.
This procedure avoids computing $\text{KD}(d_{s,f}[A], d_{s,f'}[A])$ explicitly, which would be prohibitive in large databases.


\subsection{Extending Embeddings to New Tuples}
We consider the situation where a new $R$-fact $f_\text{new}$ is
inserted into the database $D$, and our goal is to extend the existing
embedding $\varphi$ over $D$ to incorporate $f_\text{new}$. Hence, our
goal is to determine the vector $\varphi(f_\text{new})\in\mathbb{R}^{d}$. 

Let $f_\text{old}$ be an $R$-fact such that the embedding
$\varphi(f_\text{old})\in\mathbb{R}^{d}$ is already known and let $(s,A) \in \mathcal{T}(R, \ell_\text{max})$.
Then $\psi(s, A)\in\mathbb{R}^{d \times d}$ is a known matrix (that we referred to in Equation~\eqref{eq:objective}).
We wish for our new embedding $\varphi(f_\text{new})$ to satisfy the
objective in Equation~\eqref{eq:objective} with respect to
$f=f_\text{new}$ and $f'=f_\text{old}$:
\begin{equation}
    \varphi(f_\text{new})^{\top}\!\cdot\psi(s, A) \cdot \varphi(f_\text{old}) \overset{!}{=} \text{KD}(d_{s,f_\text{old}}[A], d_{s,f_\text{new}}[A])
  \end{equation}
  Hence, we obtain a linear equation $\varphi(f_\text{new})^{\top}\cdot c=y$ where $c\in\mathbb{R}^{d}$ and $y\in \mathbb{R}$ are known.
  If we stack these linear equations for many choices of $(f',s,A)$, then we obtain an overdetermined system of linear equations that we can (approximately) solve for $\varphi(f_\text{new})$.
  
  We randomly sample a sufficiently large number of such triples.
  In particular, we sample $n_\text{samples}^\text{new}\in \mathbb{N}$ distinct samples for each $(s,A) \in \mathcal{T}(R, \ell_\text{max})$ where $n_\text{samples}^\text{new}$ is a hyperparameter.
  Let $k$ be the total number of drawn samples and let $(f_i,s_i,A_i)$ be the $i$-th sample with $i\in[k]$.
  We define the matrix $C \in \mathbb{R}^{k \times d}$ such that
\begin{equation}
    C_i = \psi(s_i, A_i) \cdot \varphi(f_i).
\end{equation}
Define $b \in \mathbb{R}^{k}$ to be the vector with
\begin{equation}
    b_{i} = \text{KD}(d_{s_i,f_i}[A_i], d_{s_i,f_\text{new}}[A_i]).
\end{equation}
To obtain a new embedding for $f_\text{new}$ we solve for
\begin{equation}
    C \cdot \varphi(f_\text{new}) = b.
\end{equation}
Thus, we can infer embeddings for novel data simply by solving systems of linear equations.
Note that we aim to find an approximate solution, since there are no exact solution for overdetermined linear systems in general.
Any standard method for solving such systems can be applied.
In our case, we use the pseudoinverse $C^+$ of $C$ to obtain a solution that is optimal in the Euclidean norm:
\begin{equation}
    \varphi(f_\text{new}) = C^+ \cdot b
\end{equation}

\subsection{Hyperparameters}
The main hyperparameters are the embedding dimension $d$, the maximum walk length $\ell_\text{max}$ and the number of samples $n_\text{samples}$, as described in Section \ref{sec:opt}.
When extending a \forward embedding to new tuples $n_\text{samples}^\text{new}$ is an additional hyperparameter. 
The batch size, learning rate and number of epochs of the gradient descent training are additional parameters.
The domain kernels $\kappa_{R.A}$ for each domain can also be viewed as hyperparameters.
However, most common data types allow for simple default choices.



\section{Experimental Evaluation}\label{sec:experiments}
We now describe our experimental study. The goal of this study is to evaluate the quality of the embeddings produced, in both the static phase and (more importantly) the dynamic phase. Our quality evaluation is via a collection of downstream tasks of tuple classification. 
(We can also view the tuple-classification task as column prediction, where the class of the tuple can be seen as a new column.) 
 We focus on databases that involve multiple relations, and particularly relations that are not the target of the downstream tasks; yet these relations can be used as context for inferring embeddings, as we do in our proposed embedding \forward.  We first describe the downstream tasks (Section~\ref{sec:exp:tasks}),
the compared methods (Section~\ref{sec:methods}),
and the general runtime setup (Section~\ref{sec:exp:setup}). We then describe our experiments on the static setting (Section~\ref{sec:exp:static}) and the dynamic setting (Section~\ref{sec:exp:dynamic}). Finally we discuss the execution times of the algorithms (Section~\ref{sec:exp:times}) and the main conclusions of the experimental section (Section~\ref{sec:exp:conclusions}).

\subsection{Datasets and Tasks}\label{sec:exp:tasks}
We now describe the datasets (tasks) of our experiments. 
The  information on the structure of the datasets is summarized in Table~\ref{tab:Datasets}. 
Each dataset is a database of multiple relations, where one relation contains an attribute that we wish to predict. 
Hereafter, we refer to this relation as the \e{prediction relation}. 
Note that neither \forward nor \nodetovec see the predicted attribute during training.
\begin{table*}[t]
 \caption{\label{tab:Datasets}Information about the structure of the datasets used in the experiments.}
\centering
\begin{tabular}{|c|c|c|c|c|c|c|}
 \hline
\cc {Dataset} & \cc {Prediction Rel.} &\cc {Prediction Attr.} &\cc {\#Samples} &\cc {\#Relations} &\cc {\#Tuples} &\cc {\#Attributes} \\ \hline
{\textbf{Hepatitis}} & Dispat & type & $500$ & $7$ & $12927$ & $26$ \\ \hline
{\textbf{Genes}} & Classification & localization & $862$ & $3$ & $6063$ & $15$ \\ \hline
{\textbf{Mutagenesis}} & Molecule & mutagenic & $188$ & $3$ & $10324$ & $14$ \\ \hline
{\textbf{World}} & Country & continent & $239$ & $3$ & $5411$ & $24$ \\ \hline
{\textbf{Mondial}} & Target & target & $206$ & $40$ & $21497$ & $167$ \\ \hline
\end{tabular}
\end{table*}

\subsubsection{Hepatitis}
This database is from the 2002 ECML/PKDD Discovery Challenge.\footnote{\url{https://sorry.vse.cz/~berka/challenge/PAST/}} We use the modified version of Neville et al.~\cite{DBLP:conf/kdd/NevilleJFH03}. The goal in this task is to predict the \att{type} column, which is either \e{Hepatitis B} or \e{Hepatitis C} based on medical examinations. There are in total 206 instances of the former and 484 cases of the latter. The relation with the predicted column contains, in addition to the type classification, the age, sex and identifier of the patient.  The other relations contain the rest of the medical data. The dataset contains seven relations with a total of 26 attributes and 12,927 tuples.

\subsubsection{Mondial}
This dataset contains information from multiple geographical resources~\cite{mondial}. We predict the \att{religion} of a country.
There are $114$ countries classified as \e{Christian} and 71 as \e{non-Christian}. The prediction is based on a variety of fields such as the language, population, geography, and government of the country. The target relation, where we predict a column, is binary---it contains only the name of the country and the (predicted) classification. The dataset contains 40 different relations with a total of 167 attributes and 21,497 tuples. We use the whole database and use the \rel{Target} relation as the prediction relation as previously done by Bina et al.~\cite{DBLP:journals/dss/BinaSCQX13}.

\subsubsection{Genes}
This dataset is from the KDD 2001 competition~\cite{10.1145/507515.507523}, and contains data from genomic and drug-design applications. We predict the \att{localization} of the gene, based on biological data, with 15 different labels. The prediction relation contains only the class and an identifier for the gene, while the rest contain the biological data such as the function, gene type, cellular location and the expression correlation between different genes. The dataset contains 3 relations with a total of 15 attributes and 6,063 tuples. We remove two tuples which have a unique class to prevent split in-balances during cross-validation.

\subsubsection{Mutagenesis}
This dataset contains data on the mutagenicity of molecules on Salmonella typhimurium~\cite{doi:10.1021/jm00106a046}. We predict the mutagenicity (the \att{mutagenic} attribute) of the molecules, based on chemical properties of the molecule, with 122 positive samples and 63 negative samples. The prediction relation contains the binary class, molecule ID, and some of the chemical data, while the other relations contain more chemical data and information about the relations between the molecules. The dataset contains 3 relations with a total of 14 attributes and 10,324 tuples. Note that we do not use any external features, in contrast to some past methods for this dataset~\cite{Lodhi05ismutagenesis} that we revisit later on.

\subsubsection{World}
This dataset contains data on states and their cities. We predict the \att{continent} of a country with 7 different labels. The prediction is based on general data on the country such as population, GNP, Capital city and information on the spoken languages and cities. The dataset contains 3 relations, with a total of 24 attributes and 5,411 tuples.


\subsection{Compared Methods}\label{sec:methods}
We compare between the following alternatives.
\begin{itemize}
\item \underline{S.o.A.}: These are state-of-the-art methods that solve the multi-relational classification problem without using an embedding, such as multi-relational decision trees and forests~\cite{DBLP:journals/dss/BinaSCQX13, 10.1007/978-3-540-39917-9_5}, multi-relational Bayes nets~\cite{DBLP:conf/cidm/SchulteBCBX13} and Inductive Logic Programming (ILP)~\cite{Lodhi05ismutagenesis, 1626232}.
  This applies only to the static experiment.
\item \underline{N2V}: In the static phase, this is the \nodetovec method,  with our own implementation that is based on the original paper.
  In the dynamic phase, we use our adaption of \nodetovec described in Section~\ref{sec:n2vdynamic}
\item\underline{\forwardekshort}:
\forward embeddings, as defined in Section \ref{sec:method}.
\end{itemize}
In the next section, we discuss the implementation of the N2V and \forward variants.

In all datasets and experiments, we have a full separation between the embedding process and the downstream task.
This means that we generate the embedding independently from the task (as opposed to training for the task), and then use these embeddings as the input to a downstream classifier (that sees only the embeddings and none of the other database information).

Specifically, we train and apply an SVM classifier (Scikit-learn's SVC implementation)
 as the downstream machine-learning architecture, in both experiments. 
Performance assessment is conducted via $k$-fold cross validation with $k=10$ folds.

\subsection{Experimental Setup}\label{sec:exp:setup}

\subsubsection{Implementation}
We implemented \nodetovec on our own based on its original publication~\cite{10.1145/2939672.2939754}.
For the state-of-the-art results (S.o.A.) we used the reported numbers.
Both \nodetovec and \forward (in both static and dynamic versions) are implemented in Python using PyTorch~\cite{NEURIPS2019_9015}, Numpy~\cite{harris2020array} for the numerical operations, 
Scikit-learn~\cite{scikit-learn} for the downstream classifiers and validation, and NetworkX~\cite{SciPyProceedings_11} for the graph implementations.
The code is publicly available on GitHub\footnote{\url{https://github.com/toenshoff/DynamicDBEmbedding}}.


\begin{table}[b]
\caption{Hyperparameters in the \forward implementations. 
In the Genes dataset we use \#samples of 1,000, batch size of 10,000, and 10 epochs.
\label{table:hyparparams}}
\renewcommand{\arraystretch}{1.0}
\centering
\begin{tabular}{|c|l|c|}\hline
    \cc Alg.  &  \cc Param. & \cc Value \\\hline
    \multirow{5}{*}{\forwardek} 
    &  embedding dim. ($k$) & 100\\
    &  \#samples ($n_\text{samples}$) & 5,000\\
    & batch size & 50,000 \\
    & max walk len. ($\ell_{\max}$) & 1--3 \\
    & \#epochs & 5--10 \\
    \hline
    \multirow{7}{*}{\nodetovec}
    &  embedding dim. & 100\\
    &\#walks per node & 40\\
    &\#steps per walk & 30\\
    & context window & 5 \\
    &\#neg/\#pos samples & 20  \\
    & batch size & 40,000 \\
    & \#epochs & 10 \\\hline
\end{tabular}
\end{table}

Note that in the \forward implementation we embed only the relation that contains the tuples that we wish to classify. 
We use the default kernels in all of our experiments: Gaussian distance for numbers, and equality for all other data types.

\subsubsection{Hyperparameters}
The hyperparameters of the \forward  and \nodetovec implementations are listed in Table~\ref{table:hyparparams}. 
Note that we use different hyperparameters for the Genes dataset, 
 since the distribution of values between the relations is different compared to the rest of the datasets.
In the dynamic experiment the number of epochs for the second training phase of \nodetovec was set to 5 and $n_\text{samples}^\text{new}$ was set to 2500 for all \forward{} runs.

\subsubsection{Hardware}
We run the experiments on a server with 1 Intel i7 processor, 64 GB RAM, and an Nvidia RTX 2070 GPU. We utilize the GPU in all methods to achieve better runtimes.


\mginlinedone{This is a little bit confusing, as the previous paragraph refers first to static and the dynamic, and the next to static again. Also, we should explicitly refer to the methods in the next paragraph as "state-of-the-art". }

\subsection{Results for Static Database
Embeddings}\label{sec:exp:static}
Table~\ref{tab:Static Classification} summarizes the results for the static case. 
Also note that for each of the ten folds, we train a new embedding. Hence, to account for the randomness of both the folds and the embeddings, we report the standard deviation ($\pm x$) next to each number. Note that the embedding methods always see the full database, and the downstream classifier uses the different splits.
For the state-of-the-art (S.o.T.), we use the reported results from the publications that we mention in the table.
\newcolumntype{P}[1]{>{\centering\arraybackslash}m{#1}}

\begin{table}[b]
    \centering
    \caption{\label{tab:Static Classification}Accuracy for static classification, including standard deviation
    ($\pm$). S.o.A.~stands for state-of-the-art, where 
    we take the best result via a general (non-dataset-specific) method.}
    \label{table:static}
    {
    \begin{tabular}{|P{1.5cm}|P{1.25cm}|P{1.25cm}|P{1.25cm}|}
        \cline{1-4}
        \cc Task & \cc \forward & \cc N2V & \cc S.o.A. \\ \hline
        Hepatitis & 84.20\% $\pm4.94$ & \textbf{93.60\%} $\pm 2.5$ & 84.00\%$^{(*)}$ \newline \cite{DBLP:journals/dss/BinaSCQX13}
        \\ \hline
        Genes & \textbf{97.91}\% \newline $\pm \textbf{0.87}$ & 97.19\% \newline $\pm 1.25$ & 85.00\% \newline\cite{10.1007/978-3-540-39917-9_5} \\ \hline
        Mutagenesis & \textbf{90.00\%} $\pm 7.96$ & 88.23\% $\pm 4.56$ & \textbf{91.00\%}$^{(**)}$\newline\cite{1626232} \\ \hline
        World& 85.83\% \newline$\pm 5.34$ & \textbf{94.00\%} \newline $\pm 4.4$ & 
        77.00\% \newline\cite{1626232} \\ \hline
        Mondial & 80.95\% \newline $\pm 6.73$ & 77.62\% $\pm 5.24$ & \textbf{85\%}\newline\cite{DBLP:conf/cidm/SchulteBCBX13} \\ \hline
        \end{tabular}
        }
        \vskip1em
        \parbox{0.47\textwidth}{
    \begin{tabular}{rp{0.42\textwidth}}
    ($^{*}$)\!\!\! & 95\% achieved in a method specific for the Hepatitis dataset~\cite{afif2013ss}.\\
    ($^{**}$)\!\!\! & 96\% achieved in a method specific for the Mutagenesis dataset~\cite{Lodhi05ismutagenesis}.
    \end{tabular}
    }
\end{table}

{
\begin{figure*}[t]
  \centering
  \subfigure[Genes\label{fig:genes}]{
    \begin{tikzpicture}
\begin{axis}[
    width=7.5cm,
    height=4.5cm,
    xlabel={Ratio of new data (\%)},
    ylabel={Accuracy (\%)},
    xmin=5, xmax=95,
    ymin=0, ymax=105,
    xtick={10,20,30,40,50,60,70,80,90},
    ytick={0,20,40,60,80,100},
    xtick pos=left,
    ytick pos=left,
    y label style={at={(axis description cs:0.12,0.5)},anchor=south},
    legend style={at={(1.05,0.5)},anchor=west},
    ymajorgrids=true,
    grid style=dashed,
]

\addplot+[
    color=blue,
    mark=square,
    mark options={solid,fill=blue},
    ]
    coordinates {
    (10,96)(20,94.50)(30,92.95)(40,92)(50,89)(60,83.44)(70,75.37)(80,64.7)(90,50.75)
    };
    
\addplot[
    color=red,
    mark=o,
    ]
    coordinates {
    (10,98.49)(20,96.97)(30,96.71)(40,96.22)(50,95.26)(60,94.11)(70,92.59)(80,88.87)(90,76.66)
    };
    
\addplot[
    color=black,
    ]
    coordinates {
    (5,42.5)(95,42.5)
    };
    
\legend{Node2Vec, FoRWaRD, Baseline}
    
\end{axis}
\end{tikzpicture}
  }\\
  \subfigure[Hepatitis\label{fig:hepatitis}]{
    \begin{tikzpicture}
\begin{axis}[
    width=7.5cm,
    height=4.5cm,
    xlabel={Ratio of new data (\%)},
    ylabel={Accuracy (\%)},
    xmin=5, xmax=95,
    ymin=0, ymax=105,
    xtick={10,20,30,40,50,60,70,80,90},
    ytick={0,20,40,60,80,100},
    xtick pos=left,
    ytick pos=left,
    y label style={at={(axis description cs:0.12,0.5)},anchor=south},
    legend style={at={(1.05,0.5)},anchor=west},
    ymajorgrids=true,
    grid style=dashed,
]

\addplot[
    color=blue,
    mark=square,
    ]
    coordinates {
    (10,92.6)(20,91)(30,92.22)(40,89.65)(50,87.16)(60,85.47)(70,81.89)(80,75.45)(90,68.04)
    };
    
\addplot[
    color=red,
    mark=o,
    ]
    coordinates {
    (10,84.2)(20,84.9)(30,83.87)(40,82.75)(50,79.96)(60,79.33)(70,78.26)(80,75.92)(90,72.02)
    };
    
\addplot[
    color=black,
    ]
    coordinates {
    (5,58.8)(95,58.8)
    };
    
    
\end{axis}
\end{tikzpicture}
  }\quad\quad\quad\quad
  \subfigure[World  \label{fig:world}]{
    \begin{tikzpicture}
\begin{axis}[
    width=7.5cm,
    height=4.5cm,
    xlabel={Ratio of new data (\%)},
    ylabel={Accuracy (\%)},
    xmin=5, xmax=95,
    ymin=0, ymax=105,
    xtick={10,20,30,40,50,60,70,80,90},
    ytick={0,20,40,60,80,100},
    xtick pos=left,
    ytick pos=left,
    y label style={at={(axis description cs:0.12,0.5)},anchor=south},
    legend style={at={(1.05,0.5)},anchor=west},
    ymajorgrids=true,
    grid style=dashed,
]

\addplot[
    color=blue,
    mark=square,
    ]
    coordinates {
    (10,94.58)(20,92.08)(30,89.58)(40,90.31)(50,88.08)(60,88.96)(70,83.69)(80,76.46)(90,24)
    };
    
\addplot[
    color=red,
    mark=o,
    ]
    coordinates {
    (10,77.08)(20,75.83)(30,74.44)(40,73.44)(50,69.17)(60,65.9)(70,61.01)(80,51.35)(90,44.17)
    };
    
\addplot[
    color=black,
    ]
    coordinates {
    (5,24.2)(95,24.2)
    };
    
    
\end{axis}
\end{tikzpicture}
  }\\
   \subfigure[Mondial  \label{fig:mondial}]{
     \begin{tikzpicture}
\begin{axis}[
    width=7.5cm,
    height=4.5cm,
    xlabel={Ratio of new data (\%)},
    ylabel={Accuracy (\%)},
    xmin=5, xmax=95,
    ymin=0, ymax=105,
    xtick={10,20,30,40,50,60,70,80,90},
    ytick={0,20,40,60,80,100},
    xtick pos=left,
    ytick pos=left,
    y label style={at={(axis description cs:0.12,0.5)},anchor=south},
    legend style={at={(1.05,0.5)},anchor=west},
    ymajorgrids=true,
    grid style=dashed,
]

\addplot[
    color=blue,
    mark=square,
    ]
    coordinates {
    (10,76.67)(20,80.95)(30,84.35)(40,81.08)(50,80.49)(60,79.11)(70,77.03)(80,76.06)(90,61.40)
    };
    
\addplot[
    color=red,
    mark=o,
    ]
    coordinates {
    (10,80.47)(20,84.28)(30,87.1)(40,83.37)(50,82.43)(60,82.50)(70,82.48)(80,82.12)(90,78.76)
    };
    
\addplot[
    color=black,
    ]
    coordinates {
    (5,54.8)(95,54.8)
    };
    
    
\end{axis}
\end{tikzpicture}
     }\quad\quad\quad\quad
    \subfigure[Mutagenesis \label{fig:mutagenesis}]{
      \begin{tikzpicture}
\begin{axis}[
    width=7.5cm,
    height=4.5cm,
    xlabel={Ratio of new data (\%)},
    ylabel={Accuracy (\%)},
    xmin=5, xmax=95,
    ymin=0, ymax=105,
    xtick={10,20,30,40,50,60,70,80,90},
    ytick={0,20,40,60,80,100},
    xtick pos=left,
    ytick pos=left,
    y label style={at={(axis description cs:0.12,0.5)},anchor=south},
    legend style={at={(1.05,0.5)},anchor=west},
    ymajorgrids=true,
    grid style=dashed,
]

\addplot[
    color=blue,
    mark=square,
    ]
    coordinates {
    (10,87.89)(20,87.63)(30,87.02)(40,87.63)(50,85.53)(60,80.97)(70,78.79)(80,70.99)(90,68.18)
    };
    
\addplot[
    color=red,
    mark=o,
    ]
    coordinates {
    (10,89.47)(20,87.11)(30,87.54)(40,89.34)(50,89.04)(60,85.4)(70,84.77)(80,84.30)(90,81.59)
    };
    
\addplot[
    color=black,
    ]
    coordinates {
    (5,66.4)(95,66.4)
    };
    
    
\end{axis}
\end{tikzpicture}
      }
      \caption{Results for the dynamic experiment - baseline is the accuracy of guessing the most common class. We report the accuracy as a function of the percentage of new tuples.}
      \label{fig:dynamic_results}
\end{figure*}
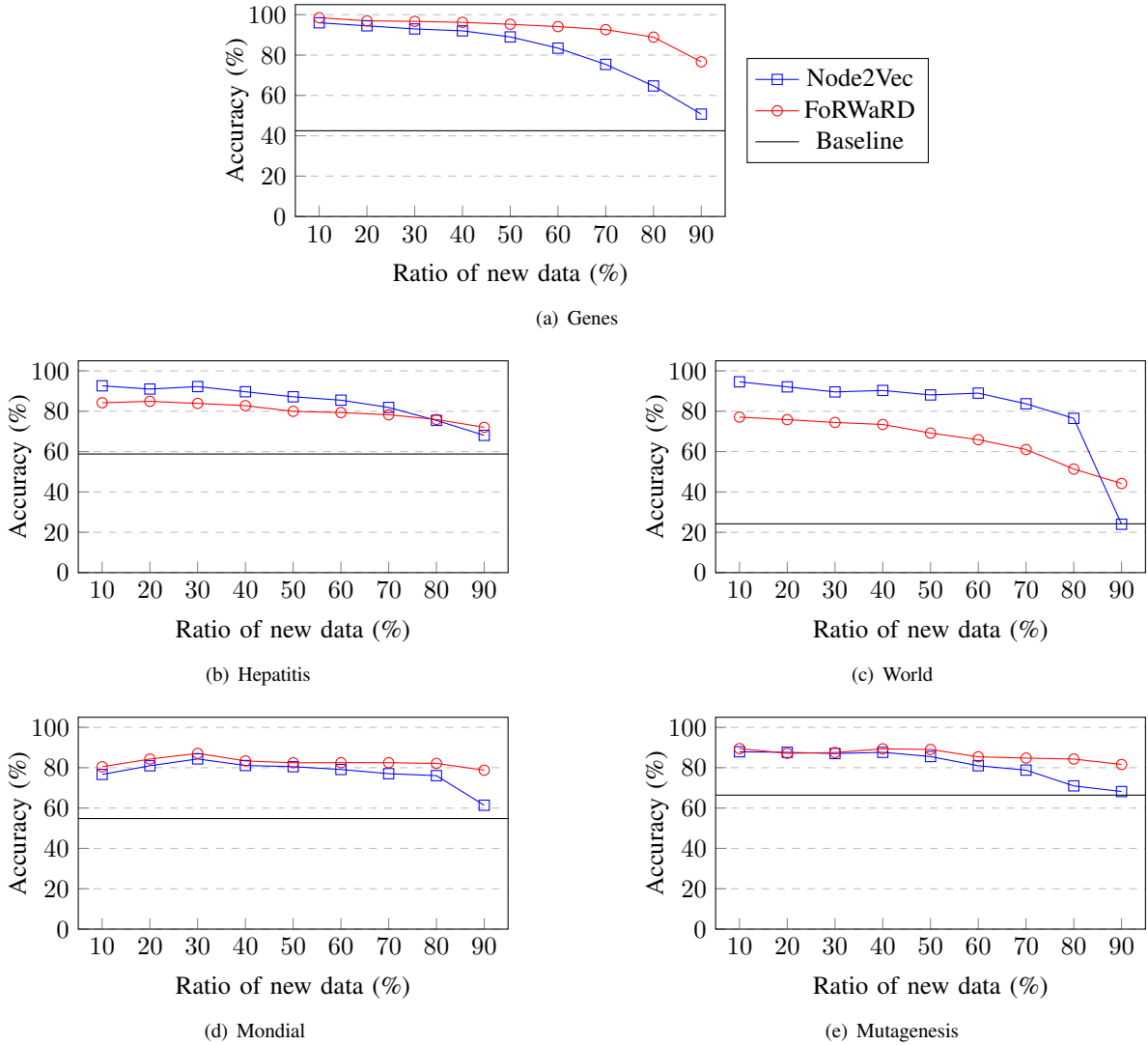
}

As can be seen in Table~\ref{tab:Static Classification}, both our methods perform better than the state-of-the-art methods on most datasets and are competitive with dataset-specific methods even without using external knowledge. Furthermore, the experiment shows that the embeddings produced with \forward are as good as the embeddings produced by the \nodetovec\ based algorithm for static classification. 

We also offer some interesting insights about the algorithms:
    First, \nodetovec preforms better on Hepatitis and World - two datasets where most of the information lays in \e{categorical} data. This is in-fact expected as the graph that is used by \nodetovec encodes this data very well.
    Second, the standard deviations of all methods are on the same scale in all datasets.
    Third, both \nodetovec and \forward excel at the Genes dataset, managing to capture the database's structure almost perfectly. Generally, both methods are competitive with the state-of-the-art in all of our datasets.
    
All in all, we can conclude that both \nodetovec and \forward achieve very good results on these datasets, and that the structure of the database affects the quality of the embeddings - and both manage to capture the structure very well.


{
\begin{table*}
\caption{\label{tab:dynamic90}Accuracy and standard deviation ($\pm$) for the dynamic experiment with a ratio of 10\% new tuples. We report results for both the all-at-once embedding extension and the one-by-one setup}
\label{table:dynamic}
\centering
    \begin{tabular}{|c|c|c|c|c|}
    \cline{1-5}
    \cc & \multicolumn{2}{c|}{\cc All at Once}  & \multicolumn{2}{c|}{\cc One by One} \\ \hline 
    \cc Task  & \cc \nodetovec & \cc \forward & \cc \nodetovec & \cc \forward \\ \hline
    Hepatitis   & $\textbf{93.34}\% \pm \textbf{2.70}$ & $82.20\% \pm 4.94$ & $\textbf{92.60}\% \pm \textbf{2.37}$ & $84.20\% \pm 5.02$ \\ \hline
    Genes       & $94.50\% \pm 1.89$ & $\textbf{97.91}\% \pm \textbf{0.87}$ & $96.20\% \pm 0.78$ & $\textbf{98.49}\% \pm \textbf{0.53}$ \\ \hline
    Mutagenesis & $87.58\% \pm 7.80$ & $\textbf{90.00}\% \pm \textbf{6.84}$ & $87.89\% \pm 7.82$ & $\textbf{89.47}\% \pm \textbf{6.66}$ \\ \hline
    World       & $\textbf{91.25}\% \pm \textbf{4.95}$ & $87.50\% \pm 3.73$ & $\textbf{94.58}\% \pm \textbf{4.58}$ & $77.08\% \pm 5.67$ \\ \hline
    Mondial     & $77.62\% \pm 6.75$ & $\textbf{80.00}\% \pm \textbf{7.32}$ & $76.67\% \pm 7.20$ & $\textbf{80.47}\% \pm \textbf{7.20}$ \\ \hline
    \end{tabular}
\end{table*}
}


\begin{table}[t]
   \caption{\label{tab:ExecutionTimesStatic}Execution times (in seconds) to compute static embeddings with \nodetovec and \forward.}
   \centering
    \begin{tabular}{|c|c|c|}
    \cline{1-3}
    \cc Task & \cc {\nodetovec} & \cc {\forward} \\ \hline
    Hepatitis & $\textbf{189}$ & $540$ \\ \hline
    Genes & $\textbf{78}$ & $204$ \\ \hline
    Mutagenesis & $\textbf{166}$ & $230$ \\ \hline
    World & $\textbf{219}$ & $440$ \\ \hline
    Mondial & $\textbf{462}$ & $810$ \\ \hline
    \end{tabular}
\bigskip   
\end{table}

\subsection{Results for Dynamic Database Embeddings}\label{sec:exp:dynamic}
\subsubsection{Experiment Description}
In this experiment we aim to measure how well we can produce new embeddings when new data arrives.
We test that by splitting the database into two parts and treat the first part as the static database, and the other part as the new data that arrives.
The experiment is comprised of five steps:
\begin{enumerate}
    \item Partition all facts of the database into two sets $\mathcal{F}_\text{old}$ and $\mathcal{F}_\text{new}$.
    The facts in $\mathcal{F}_\text{new}$ are then removed from the database.
    \item Train an embedding \e{only} on the static part of the split.
    \item Label the vectors in this embedding using the correct class labels and train a downstream classifier on these labelled data.
    \item Add the facts from from $\mathcal{F}_\text{new}$ back to the database to simulate the arrival of new data. Generate new embeddings for these new facts.
    \item Evaluate the trained classifier only on the embeddings of ``new'' data.
\end{enumerate}
It is important to note that, by (5), the accuracy results we obtain refer exclusively to new data added after the embedding was trained.

The first question that arises is how to split the data in a way that simulates a real-life scenario.
In real-world scenarios, new facts tend to arrive in batches that are distributed across all relations. 
For our example database from Figure~\ref{fig:dbexample}, a new fact that is added to the \rel{Movies} relation may be accompanied by new actors or new collaborations that are also added to the database.
In our experiments we would like to simulate this behavior, where the new facts that arrive are are semantically related. 
Note that such a setup is more challenging than a simple random partition, since the new facts are less connected to the old data.

Formally, we compute the partition as follows.
We first partition the prediction table according to a specified ratio of \emph{old} and \emph{new} tuples.
This split is randomly chosen and stratified, i.e.\ the classes of the downstream task are (roughly) equally distributed in both partitions.
We then iteratively remove the \emph{new} tuples from the prediction relation in a random order.
We remove each tuple with an ``On Delete Cascade'' deletion, which will automatically fix the foreign-key constraints throughout the database.
In particular, data that is only referenced by the tuple that is being deleted is also removed from the database. 
The tuples that remain after these deletions form $\mathcal{F}_\text{old}$ while the deleted tuples form $\mathcal{F}_\text{new}$.

\begin{example}
Consider again the database of Figure~\ref{fig:dbexample} and suppose that our prediction relation is \rel{Collaborations}. If we remove the fact $c_1$, then we will also remove the fact $m_4$ (Interstellar) from the \rel{Movies} relation, and the tuple $a_2$ (Watanabe) from the \rel{Actors} relation. Note that we will \e{not} remove $a_1$ (DiCaprio) as it is still connected to non-removed facts in other relations (specifically, it is connected to $c_4$ in the \rel{Collaborations} relation).
\qed
\end{example}


After computing the initial embedding of $\mathcal{F}_\text{old}$ we add the facts of $\mathcal{F}_\text{new}$ back to the database and extend the embedding to this new data.
By default, we add the deleted facts from the prediction table \emph{one-by-one} in the inverse order of their deletion.
With each new tuple in the prediction table we also add all the referenced tuples that were removed in the ``On Delete Cascade'' deletion. 
We then extend the embedding to these new tuples before the next tuple is added to the prediction table.
Note that in this setting we do not recompute the paths starting at the old tuples, neither for \nodetovec{} nor for \forward{}.
The distribution of paths originating at the old tuples is changing as new data is added.
However, recomputing these paths for each new tuple would slow down both methods substantially.

Additionally, we also study the case where all new tuples are added at once.
We refer to this as a \emph{all-at-once} embedding extension.
In this setting we do recompute paths in the old data, since this only needs to be done once.


\begin{table*}
  \caption{Average time (in seconds) to embed (training and inferring) a new tuple for \nodetovec and \forward.}
    \label{tab:ExecutionTimesDynamic}
    \centering
    \begin{tabular}{|c|c|c|c|c|}
    \hline
    \cc & \multicolumn{2}{c|}{\cc All at Once}  & \multicolumn{2}{c|}{\cc One by One} \\ \hline 
    \cc Task & \cc \nodetovec & \cc \forward & \cc \nodetovec & \cc \forward \\ \hline
    Hepatitis   & $\textbf{0.265}$ & $0.620$  & $0.679$ & $\textbf{0.111}$ \\ \hline
    Genes       & $\textbf{0.062}$ & $0.176$ & $0.173$ & $\textbf{0.079}$ \\ \hline
    Mutagenesis & $0.650$ & $\textbf{0.280}$ & $0.764$ & $\textbf{0.134}$ \\ \hline
    World       & $\textbf{0.640}$ & $0.733$ & $0.283$ & $\textbf{0.149}$ \\ \hline
    Mondial     & $1.550$ & $\textbf{1.090}$ & $1.710$ & $\textbf{0.385}$ \\ \hline
    \end{tabular}    
\end{table*}

In our experiment we vary the ratio of old and new facts to understand how the relative amount of data effects the quality of the induced embeddings for the new data.
For each tested ratio, we run the experiment 10 times with different partitions of the database.
We measure the evaluation accuracy of the downstream classifier on the new data and the corresponding standard deviation across the 10 runs.
\subsubsection{Results}

Figure~\ref{fig:dynamic_results} provides the accuracies achieved in the one-by-one extension of the embeddings. 
We omit the results of the all-at-once extension in this figure, as the accuracy results are very similar in both versions.
The term ``baseline'' here means the accuracy obtained by always predicting the most common class.
Both \forward and \nodetovec have similar performance and are able to produce useful embeddings for new tuples. 
As the percentage of new data increases the performance of both methods tends to decrease.
However, this decrease is fairly slow and only becomes more significant when over 50\% of the data is removed.

Table~\ref{tab:dynamic90} compares the two methods in both experimental setups (all-at-once vs one-by-one).
Here, we look at the specific scenario where 10\% of tuples are new data.
Recall that in the all-at-once setup we do recompute paths for all tuples prior to extending the embedding to the new data.
This is not done in the one-by-one experiment.
One would expect the performance of both methods to be better in the all-at-once setup, since it takes into account paths which start at old tuples and traverse new data.
This information is otherwise neglected.
Surprisingly, the results are very similar in both setups.
Only on the World dataset \forward{} achieves a significantly lower accuracy in the one-by-one setting when compared to the all-at-once embedding extension.
Otherwise, the results are only marginally worse or even slightly better when adding tuples one-by-one without recomputing paths for old data.


Overall, we observe that both \forward and \nodetovec empirically achieve the goal of producing viable embeddings for new data, while keeping the old embeddings intact.
In most of the datasets, the accuracy is almost as good as the static case, even when up to 50\% of the data is new. 
\subsection{Execution Times}\label{sec:exp:times}
Here, we compare \forward's\ execution times to that of \nodetovec in both the static and dynamic phases.
Table~\ref{tab:ExecutionTimesStatic} reports the runtime for the static classification.
We observe that \nodetovec\ is faster than \forward on all of the datasets. 


Finally, we are also interested in the average time for generating an embedding for a newly arrived tuple, looking at both versions of the dynamic experiment.
The results are shown in Table~\ref{tab:ExecutionTimesDynamic}. 
The reported numbers are the average time (in seconds) that it takes to embed a newly-arrived tuple - we already have the embedding for the old tuples, and we measure only how much time it takes to train and infer a new embedding for the new tuple.
Note that these numbers change dramatically from one dataset to another, as according to the experiment each new tuple in the classification relation is also accompanied by new tuples from other relations, thus the structure of the database affects the execution time.
In the all-at-once setting, neither method is consistently faster than the other.
In the `one-by-one setting, where new tuples arrive iteratively, \forward is significantly faster than \nodetovec across all datasets. 
Note that this is expected, because for \nodetovec we have to re-train the embedding using gradient descent for every new tuple that arrives, where as for \forward we only have to solve a system of linear equations.
This insight was essential in the design of \forward.


\balance



\subsection{Conclusions}\label{sec:exp:conclusions}
We offer some final conclusions regarding the experiments.

First, the tested methods all provide good tuple embeddings in the static setting, across all datasets, competing with the state-of-the-art methods.

Second, in the dynamic phase both \nodetovec and \forward give good embeddings even when more than half of the data is new.
    

Finally, which method performs best strongly depends on the database. Both \nodetovec and \forward can significantly outperform the other method respectively.

\section{Concluding Remarks}\label{sec:conclusions}

We studied the stable database embedding problem, that is, the problem of computing vector embeddings of dynamically changing databases where the embedding of tuples remains stable once they have been added to the database. 
Past techniques for the static database embedding
problem are not designed for dynamic database embedding. We propose two algorithmic solutions to this problem, an adaptation of the node embedding algorithm \nodetovec for graphs and a new algorithm forward \forward. Both of our embedding methods incorporate the structure of the entire databases with relations linked by foreign key constraints, while previous embedding methods focus on embedding single relations. Our experiments with a variety of downstream tasks demonstrate that both methods perform surprisingly well even as large parts of the database have changes in the dynamic process, with a slight edge for the \forward method in the dynamic setting.

In future work, it will be important to study the performance of our embedding algorithm with respect to downstream machine-learning tasks other than column
prediction: record
linking~\cite{DBLP:conf/sigmod/MudgalLRDPKDAR18,DBLP:journals/pvldb/EbraheemTJOT18},
entity
resolution~\cite{DBLP:conf/sigmod/CappuzzoPT20,DBLP:conf/aistats/KushagraBI19},
data
imputation~\cite{DBLP:conf/mlsys/WuZIR20,DBLP:journals/apin/LakshminarayanHS99},
data
cleaning~\cite{DBLP:conf/mlsys/WuZIR20,DBLP:conf/icdt/SaIKRR19,DBLP:journals/pvldb/AbedjanCDFIOPST16}
and so on.

\mginlinedone{New paragraph on deletions and privacy}
We argued in the Introduction that tuple deletion is a trivial operation in our dynamic setup where we preserve the embedding of existing tuples. This is why we focused on tuple insertions. However, there is a subtle issue about deletions that poses interesting questions somewhat orthogonal to what we study here. When deleting a tuple $t^-$ and the corresponding point from the embedding, we do not delete all information about $t^-$, since the existence of the tuple had impact on how the other tuples were embedded. This has consequences for privacy considerations. For instance, if a user wants to be deleted from a database, then all information about the user must be deleted; this may even be a legal requirement.
For this, the embedding of the remaining tuples needs to be adapted. Rather than re-computing the whole embedding, we may try to find a minimal set of changes that removes all information about the deleted tuple, but keeps the overall embedding intact.


\bibliographystyle{acm}
\bibliography{bibliography}
\end{document}